\documentclass[preprint,preprintnumbers,amsmath,amssymb]{revtex4}
\usepackage{amsmath}
\usepackage{graphicx}

\begin{document}
\title{The parity of specular Andreev reflection under mirror operation in zigzag graphene ribbon}

\author{Yanxia Xing$^{1,3}$, Jian Wang$^{1,\ast}$, and Qing-feng Sun$^{2,\dag}$}

\affiliation{$^1$Department of Physics and the Center of Theoretical
and Computational Physics, The University of Hong Kong, Pokfulam
Road, Hong Kong, China. \\
$^2$Beijing National Lab for Condensed Matter Physics and Institute
of Physics, Chinese Academy of Sciences, Beijing 100080, China. \\
$^3$Department of Physics, Beijing Institute of Technology, Beijing
100081, China.\\}

\begin{abstract}
It is known that the parity of reflection amplitude can either be
even or odd under the mirror operation. Up to now, all the parities
of reflection amplitude in the one-mode energy region are even under
the mirror operation. In this paper, we give an example of odd
parity for Andreev reflection (AR) in a three-terminal
graphene-supercondutor hybrid systems. We found that the parity is
even for the Andreev retroreflection (ARR) and odd for specular
Andreev reflection (SAR). We attribute this remarkable phenomenon to
the distinct topology of the band structure of graphene and the
specular Andreev reflection involving two energy bands with
different parity symmetry. As a result of odd parity of SAR, the SAR
probability of a four-terminal system with two superconducting leads
(two reflection interfaces) can be zero even when the system is
asymmetric due to the quantum interference of two ARs.
\end{abstract}

\pacs{72.80.Vp,
74.45.+c,
73.40.-c,
74.25.F-
}\maketitle


Since the experimental realization of graphene\cite{ref1}, it has
become an exciting arena for theoretical and technological
investigations.\cite{ref2} A number of new phenomena have been
predicted and verified experimentally. For instance, in the presence
of magnetic field, it exhibits a distinctive half-integer quantum
Hall effect.\cite{ref1} Its quasi-particles obey the Dirac-like
equation and have relativistic-like behaviors.\cite{ref2} Due to the
relativistic effect, the Klein tunneling occurs where an incident
electron in graphene can pass through a potential barrier with
probability one.\cite{Klein} Then a graphene p-n junction can be
used to focus Dirac electron current with a negative refractive
index.\cite{altshuler,lens}

Since good contacts between superconducting leads and graphene have
been realized experimentally,\cite{ref16} the transport study
through graphene based normal-metal-superconductor (GNS)
heterojunction becomes feasible. In the presence of a normal metal
(graphene)-superconducting interface, an incoming electron converts
into a hole and a cooper pair is formed that enters the
superconductor. Due to the relativistic nature of the electron in
graphene, the electron-hole conversion can either be intraband
(within conduction or valence band) or interband (between conduction
and valence bands). When the electron-hole conversion is intraband,
it corresponds to the usual Andreev reflection (AR)\cite{ref15} or
Andreev retroreflection (ARR) because the reflected hole is along
the incident direction. This ARR occurs for both relativistic and
non-relativistic electrons. When the electron-hole conversion is
interband, the reflected hole is along specular direction and a
specular Andreev reflection (SAR) takes place,\cite{Beenakker} which
can lead to novel phenomena as we will discuss below.

It is known that the parity is a fundamental quantity in physics and
reflection is a general physical phenomenon in nature. In this
paper, we discuss the parity of reflection amplitude for graphene in
contact with superconductor leads. In general, the parity of a
reflection amplitude can be either even or odd when the system is
under mirror operation. However, for all previous known reflection
events, the reflection amplitudes in the one-mode energy region have
even parity under the mirror operation. It is yet to find an
odd-parity reflection event. In this paper, we found, for the first
time, that the SAR amplitude has an odd parity under the mirror
operation for zigzag graphene ribbons with even number of chains.
This means that the phases of SAR amplitude for a
graphene-superconductor hybrid system and its mirror system differ
by $\pi$. We attribute this phenomenon to the unique band structure
of the graphene. Obviously this phase difference does not affect any
observable quantities for each system. When two systems couple
together, however, this $\pi$ phase manifest through quantum
interference between two SARs. So this $\pi$ phase shift has
important consequences for a four terminal device with two
superconducting leads (see Fig.2(a)). When two superconducting leads
are symmetrically attached to the device, the quantum interference
of the left and right SAR leads to a destructive or constructive
interference depending on whether the phase difference of
superconducting leads is zero or $\pi$. Importantly, when two
superconducting leads are {\it asymmetrically} attached to the
device, the same interference pattern occurs provided that the Dirac
point $E_0$ is in line with the condensate of superconducting lead.
The quantum interference between pairs of the AR can be tuned by
shifting the Dirac point, the asymmetry of the two superconducting
leads, as well as the phase between two superconducting leads. Due
to the odd parity of SAR, the interference pattern for SAR is phase
contrasted to that of ARR where the parity is even.


Before doing numerical calculation, we first prove that the phases
of SAR amplitude of two systems (i) and (ii) in Fig.1(a) differ by
$\pi$, i.e., the parity of SAR is odd under mirror operation. Note
that for graphene systems electrons in valence and conduction band
are usually referred as electrons and holes, respectively. In the
presence of superconducting lead the reference point of electrons
and holes is the Fermi level in the superconducting lead. In the
following, we will refer electrons (holes) as electrons above
(below) Fermi level in superconducting lead. Denote $\psi^{+}_c$
($\psi_v^+$) the wavefunction of electrons in conduction (valence)
band moving in +y direction and $\psi^{-}_c$ ($\psi_v^-$) in -y
direction in the zigzag graphene nanoribbon {\it lead}. It was known
that under reflection ${\hat P}: x \rightarrow -x$, $\psi_c^{\pm}$
is symmetric while $\psi_v^{\pm}$ is anti-symmetric if the energy of
electron is in the first transmission channel\cite{select}[see
Fig.1(a)], i.e.,
\begin{eqnarray}
{\hat P} \psi_c^\pm(x,y)&=&\psi_c^\pm(-x,y) \nonumber \\
{\hat P} \psi_v^\pm(x,y)&=&-\psi_v^\pm(-x,y) .\label{mirror}
\end{eqnarray}
which is one of the unique features of zigzag edge nanoribbons with
even number of chains. Assuming the incident electron from the
terminal-1, the wavefunctions for SAR $\psi_{1,3}$ in zigzag
nanoribbon lead 1 or 3 of the system (i) can be written as
\begin{eqnarray}
\psi_1^{(i)} &=& \psi_e^+ + r_{11} \psi_e^- + r_{11A} \psi_h^-
\nonumber \\ \psi_3^{(i)} &=& t_{13} \psi_e^+ + r_{13A} \psi_h^+
\label{systema}
\end{eqnarray}
where $r_{11}$ is the normal reflection amplitude, $t_{13}$ is the
transmission amplitude, $r_{11A}$ and $r_{13A}$ are the Andreev
reflection amplitudes with the reflected hole to the terminal-1 and
3, respectively. Similarly the wavefunctions for the system (ii) are
given by
\begin{eqnarray}
\psi_1^{(ii)} &=& \psi_e^+ + {\bar r}_{11} \psi_e^- +
{\bar r}_{11A} \psi_h^- \nonumber \\
\psi_3^{(ii)} &=& {\bar t}_{13} \psi_e^+ + {\bar r}_{13A} \psi_h^+
\label{systemb}
\end{eqnarray}
Since the system (i) is related to (ii) by the reflection operator
${\hat P}$, we have $\psi_\alpha^{(i)}={\hat P} \psi_\alpha^{(ii)}$
with $\alpha=1,3$. Note that for SAR, the electron is in the
conduction band while the hole is in the valence band, i.e.,
$\psi_e=\psi_c$ and $\psi_h=\psi_v$. From this relation together
with Eqs.(\ref{mirror}), (\ref{systema}), and (\ref{systemb}), we
obtain
\begin{eqnarray}
&&r_{11A} = -{\bar r}_{11A}, ~~~ r_{13A} = - {\bar r}_{13A}\nonumber \\
&&r_{11} = {\bar r}_{11}, ~~~ t_{13} = {\bar t}_{13}
\end{eqnarray}
Note that the origin of this $\pi$ phase shift (odd parity) is the
interband conversion from the electron to the hole. Therefore the
$\pi$ phase shift does not occur for ARR since it involves only
intraband conversion. Now we verify this statement numerically using
a tight-binding model (see below for detailed description of the
model and numerical procedure). The numerical results of AR
probability $R_{11A(13A)}=|r_{11A(13A)}|^2$ for two systems are
shown in Fig.1(b). As expected the AR probability are exactly the
same for two systems. However, the phase of AR amplitudes
$r_{11A(13A)}$ denoted as $\Phi^{i,ii}_{11(13)}$ are different. It
is shown in Fig.1(c) and Fig.1(d) that ARR amplitudes
($|E_0|>|E_F|$, with $|E_F|=0.5$) are the same for two systems in
Fig.1(a) while the SAR amplitudes ($|E_0|<|E_F|$) have a $\pi$ phase
shift. It confirms the odd parity for interband electron-hole
conversion, which comes from the distinct topology of the band
structure of graphene.

To see the consequence of the odd parity of SAR, we examine a
symmetric four-terminal device with two superconducting leads
depicted in Fig.2(a) (by setting asymmetry $\delta N=0$ and phase
difference $\delta \phi=0$). For this system, two beams from
terminal-1 has a $\pi$ phase shift due to odd parity of SAR and
interferes destructively at terminal-3 giving rise to a vanishing
SAR coefficient. However, we can arrive the same conclusion using
symmetry argument as follows. Since the system is symmetric with
respect to $x=0$, we must have $r_{13A} = {\bar r}_{13A}$ when the
reflection operation along x-direction is applied. While from
Eq.(4), $r_{13A} = -{\bar r}_{13A}$. So the AR probability
$R_{13A}=|r_{13A}|^2$ for SAR can also be zero from symmetry point
of view.\cite{cheng} Therefore we conclude that the symmetric device
can not be used to test the odd parity of SAR. In the following, we
demonstrate that due to the $\pi$ phase shift the destructive
interference still occurs in a four-probe devices with two
superconducting leads attached asymmetrically and hence can be used
to test the odd parity of SAR.

For this purpose, we consider an asymmetric four-terminal device
consisting of a zigzag graphene ribbon with two superconducting
leads as shown in Fig.2(a). The Hamiltonian of the graphene
is\cite{ref19}
 $H_{0} = \sum_{\bf i} \epsilon_{\bf i} a^{\dagger}_{\bf i} a_{\bf i}
      -\sum_{<{\bf ij}>} t  a_{\bf i}^{\dagger} a_{\bf
      j}$.
Here $a_{\bf i}$ and $a_{\bf i}^{\dagger}$ are the annihilation and
creation operators at site ${\bf i}$, $\epsilon_{\bf i}$ is the
on-site energy which can be controlled experimentally by the gate
voltage\cite{ref1}, and the hopping constant $t=2.75eV$ represents
the nearest carbon bond energy. The pair potential (energy gap) of
superconducting terminal-$\beta$ with $\beta=2,4$ is
$\tilde{\Delta}_\beta=\Delta_\beta e^{i\varphi_\beta}$ with
$\Delta_{2}=\Delta_{4}=\Delta\simeq 1meV$. In numerical
calculations,\cite{cheng} we fix Fermi energy $E_F$ and tune the
Dirac point $E_0$. We have used $\Delta$ as the energy unit.


Now we study the interference between two ARs from GNS junctions as
shown in Fig.2(a) in which two superconducting leads 2 and 4 are
asymmetrically attached to the zigzag nanoribbon. The horizontal
distance $\delta N$ between two GNS junctions measures the asymmetry
of two GNS junctions. The scattering process can be qualitatively
understood as follows. For simplicity, we assume $\phi_2=\phi_4$ for
the moment. As shown schematically in Fig.2(a), for SAR the
particle-like electrons in terminal-1 split into two beams and are
scattered separately by two GNS junctions (green horizontal lines)
as holes that finally recombine at terminal-3. We examine the total
phase accumulated for each beam that involves the following three
processes. Before reaching the first GNS junction (denoted by the
left vertical green line) two beams of electrons propagate with the
same momentum $k_x$. After reaching the second GNS junction (denoted
by the right vertical green line) two beams of holes also propagate
with the same momentum $k'_x$. Obviously phases accumulated in the
above two processes for both beams are the same. Between them two
beams propagate with different momenta $k_x$ and $k'_x$. Hence the
phase difference between two beams is $\phi=(k_x-k'_x) \delta x$
with $\delta x=b\delta N$, where $b=\sqrt{3}a$ and $a$ the lattice
constant. This phase difference can be tuned by varying the Dirac
point $E_0$ or the asymmetry $\delta N$ giving rise to a complicated
interference pattern (see Fig.2). In particular, this phase
difference can be zero if $(k_x-k'_x)=0$ (i.e.,$E_0=0$) or $\delta
N=0$. In general, the total phase difference is $\phi=(k_x-k'_x)
\delta x +\phi_2-\phi_4$.


Interference pattern of AR probability $R_{13A}$ for system depicted
in Fig.2(a) with pair potential phase difference of two
superconductors $\delta\varphi=0$ and $\pi$
($\delta\varphi\equiv\varphi_2-\varphi_4$) are then plotted in
Fig.2(b) and (c), respectively. For Fig.2(b) following observations
are in order: (1) For the geometrically symmetric system ($\delta
N=0$), the interference is always destructive with zero $R_{13A}$ as
long as $|E_0|<|E_F|$.\cite{cheng} Clearly this is due to the $\pi$
phase shift depicted in Fig.1(d) and is consistent with the band
selection rule.\cite{select} (2) When Dirac point $E_0$ is in line
with the condensate energy of the superconductor, i.e., when
$E_0=0$, $R_{13A}$ is again zero no matter what value $\delta N$
assumes. This means that there is a completely destructive
interference between two beams scattered by two GNS junctions
attached asymmetrically to the graphene nano-ribbon. This behavior
can be understood as follows. When $E_0=0$ the incoming electron and
reflected hole have the same propagating momentum $k_x$ and thus
path 1 and 2 in Fig.2(a) experience the same quantum phase $k_x
\delta x$ except at the superconducting leads. Hence the total phase
difference is only due to the $\pi$ phase shift between two SARs.
(3) $R_{13A}$ is an even function of Dirac point $E_0$ because of
the electron-hole symmetry in graphene. Due to the geometric
symmetry, $R_{13A}$ is also an even function of asymmetry $\delta
N$. (4) For nonzero $E_F$, the closer the Dirac point $E_0$ to
$E_F$, the more rapidly $R_{13A}$ oscillates as we vary $\delta N$.
This is because the difference of propagating momentum $k_x-k'_x$
increases monotonically as $E_0$ approaches to $E_F$. (5) When $E_0$
is in the vicinity of $E_F$, $R_{13A}$ can reach 0.9 which is much
larger than that when $|E_0|>|E_F|$. This is because when $E_F$ is
very close to $E_0$, the edge states of zigzag ribbon begin to
contribute, then electron is easier to be scattered by two GNS
junctions located also at edges of zigzag ribbon. Considering the
pseudo-spin conservation, large $R_{13A}$ is always found in the
region of $|E_0|<|E_F|$, i.e., the SAR region. (6) There is an
overall fine oscillation with a period of $\delta N=3b$. Similar
behavior was also found in zigzag ribbons with a p-n junction where
the conductance is determined by the relative displacement $\delta$
along the p-n junction.\cite{Beenakker1} In Fig.2(c) with the
superconducting phase difference $\delta\varphi=\pi$, we see that
the interference pattern is contrary to $\delta\varphi=0$ [Fig.2(b)]
where the constructive interference becomes destructive and vice
versa.

To further analyze the interference pattern, we plot in Fig.3(a) the
total $R_{13A}$ vs Dirac point $E_0$ for different asymmetry $\delta
N$ with the phase difference between two superconducting leads
$\delta \varphi=0$ [main panel of Fig.3(a)] or $\delta \phi=\pi$
[inset of Fig.3(a)]. Clearly the interference (oscillatory) pattern
occurs only for asymmetric systems ($\delta N\neq0$) with
oscillation frequency proportional to $\delta N$. When pair
potential phase difference $\delta\varphi=\pi$ is introduced, the
interference pattern reverses, and $R_{13A}$ with $\delta N=0$
becomes the envelop function of $R_{13A}$ for all nonzero $\delta
N$. In Fig.3(b) we plot $R_{13A}$ vs $\delta N$ for different widths
$W$ of nanoribbon. It is shown clearly that $R_{13A}$ is a periodic
function of $\delta N$ with larger periodicity for larger $W$. In
the inset of Fig.3(b) we plot this period versus the width for
different $E_0$. The period $P$ is obtained in two ways: (1). from
the expression $P=2\pi/(k_x-k'_x)$ where the momenta $k_x$ and
$k_{x}'$ can be obtained from the band structure for a given $E_0$
(black symbols). (2). directly from main panel of Fig.3(b) (red
solid circle). From the inset, it clearly shows that two periods are
exactly the same giving strong evidence that the interference
pattern of AR probability are indeed from two reflected hole beams.

Finally, the interference pattern of AR probability $R_{11A}$ is
also studied (not shown). We found that only ARR probability
$R_{11A}$ ($|E_0|>E_F=0.2\Delta$) exhibits interference pattern. We
note that since there is no $\pi$ phase shift involved in ARR, when
$\delta N=0$ reflected electrons through two GNS junctions interfere
constructively when $\delta \varphi=0$ and destructively when
$\delta \varphi=\pi$ which is in contrast to SAR in Fig.2. In fact,
interference patterns of SAR and ARR are always phase contrast not
only for $\delta N=0$ but also for all other $\delta N$.

To test the odd parity of SAR experimentally, it relies on the
fabrication of high quality zigzag graphene nanoribbons. It has been
achieved by several laboratories using different methods last year
including the method to unzip the multi-walled carbon nanotube
(CNT),\cite{ref13} the anisotropic etching by thermally activated
nickel nanoparticles,\cite{ref14} and use reconstruction of the edge
to make zigzag graphene nanoribbons.\cite{experiment} In view of the
above experimental breakthrough, we expect that the setup to test
our predicted phenomenon can be realized experimentally.

To reduce the experimental challenge, we have considered an unzipped
CNT device, i.e., (n,n) CNT-zigzag graphene-(n,n) CNT, obtained by
unzipping a few unit cells in the central part of an armchair CNT
which has been achieved experimentally.\cite{ref13} For this system,
the wavefunction in the armchair CNT has the same symmetry as that
of the zigzag graphene ribbon. Following the same procedure leading
to Eq.(4), we have shown that the unzipped CNT in contact with a
superconducting lead has the odd parity under mirror operation.
Similar conclusions drawn from GNS can be obtained for unzipped CNT
with two superconducting leads.

In conclusion, up to now, the parity of reflection amplitude was
found to be even under the mirror operation. Here we have provided
an example of odd parity for the reflection amplitude, the SAR
amplitude in the zigzag graphene-superconductor hybrid system. This
odd parity is due to the combination of unique band structure of the
graphene and the electron-hole conversion involving two energy bands
with different parity symmetry. The signature of odd parity of SAR
can be found from the quantum constructive interference in a four
terminal system with two superconducting leads attached
asymmetrically. Furthermore, the interference pattern due to odd
parity of SAR is phase contrasted to that of ARR where the parity is
even.

{\bf Acknowledgement} We gratefully acknowledge the financial
support from a RGC grant (HKU 705409P) from the Government of HKSAR
and from NSF-China under Grant Nos.10974236 and 10821403.\\

\begin{figure}
\includegraphics{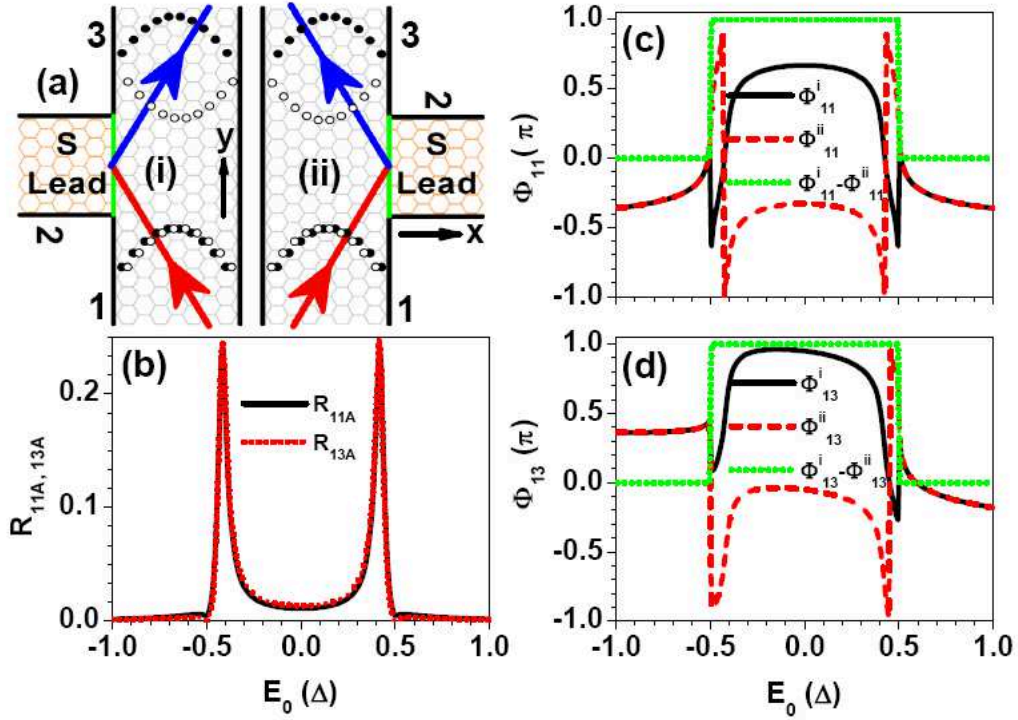}
\caption{ (Color online) Panel (a): zigzag ribbons with even number
of chains (gray honeycomb) attached by a superconducting lead on the
left and right (orange honey comb), respectively. For SAR the
incoming electrons (red arrow) are scattered by the GNS junction
(green solid line) as holes (blue arrow). The corresponding wave
functions at sublattice ``A" (solid circle) and ``B" (hollow circle)
for the lowest subband in conduction band (bottom) and the highest
subband in valence band (top) are shown schematically. Panel (b): AR
probability from terminal-1 to terminal-1 $R_{11A}$ and to
terminal-3 $R_{13A}$ vs. Dirac point $E_0$. Panels (c) and (d): AR
phase $\Phi^{i,ii}_{11}$ (c) and $\Phi^{i,ii}_{13}$ (d) of two
systems in panel (a) and their phase different
$\Phi^i_{11(13)}-\Phi^{ii}_{11(13)}$ vs. $E_0$.} \label{Fig1}
\end{figure}

\begin{figure}
\includegraphics{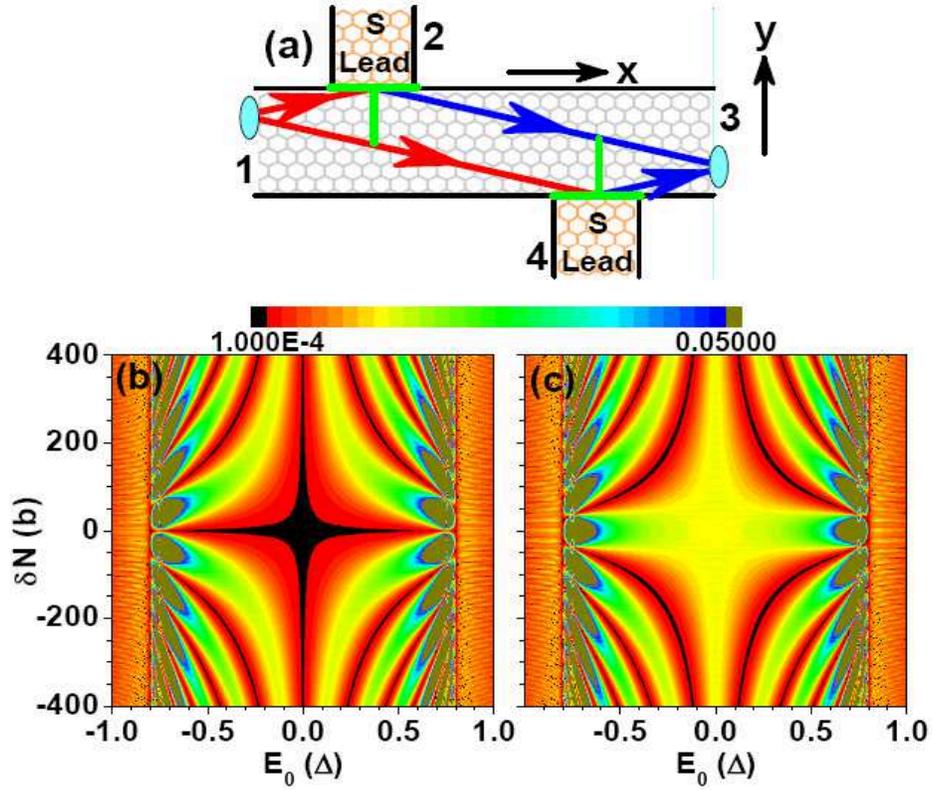}
\caption{ (Color online) Panel (a): sketch of AR interferometer in
which the zigzag ribbon is asymmetrically attached by two
superconductor lead-2 and 4. Electrons in terminal-1 can be Andreev
reflected into terminal-3 by either top or bottom GNS junction
(horizontal green lines). Panel(b) and (c): the contour plot of
$R_{13A}$ vs Dirac point $E_0$ and asymmetry $\delta N$. The phase
difference of two superconductor leads $\delta\varphi$ is zero in
panel (b) and $\pi$ in panel (c). The other parameters: Fermi energy
$E_f=0.8$, number of chains in zigzag ribbon $N=40$ corresponding to
width $60a$, the width of superconductor lead $W_S=10b$, where
$b=\sqrt{3}a$.} \label{Fig2}
\end{figure}

\begin{figure}
\includegraphics[width=14cm,totalheight=17cm, clip=]{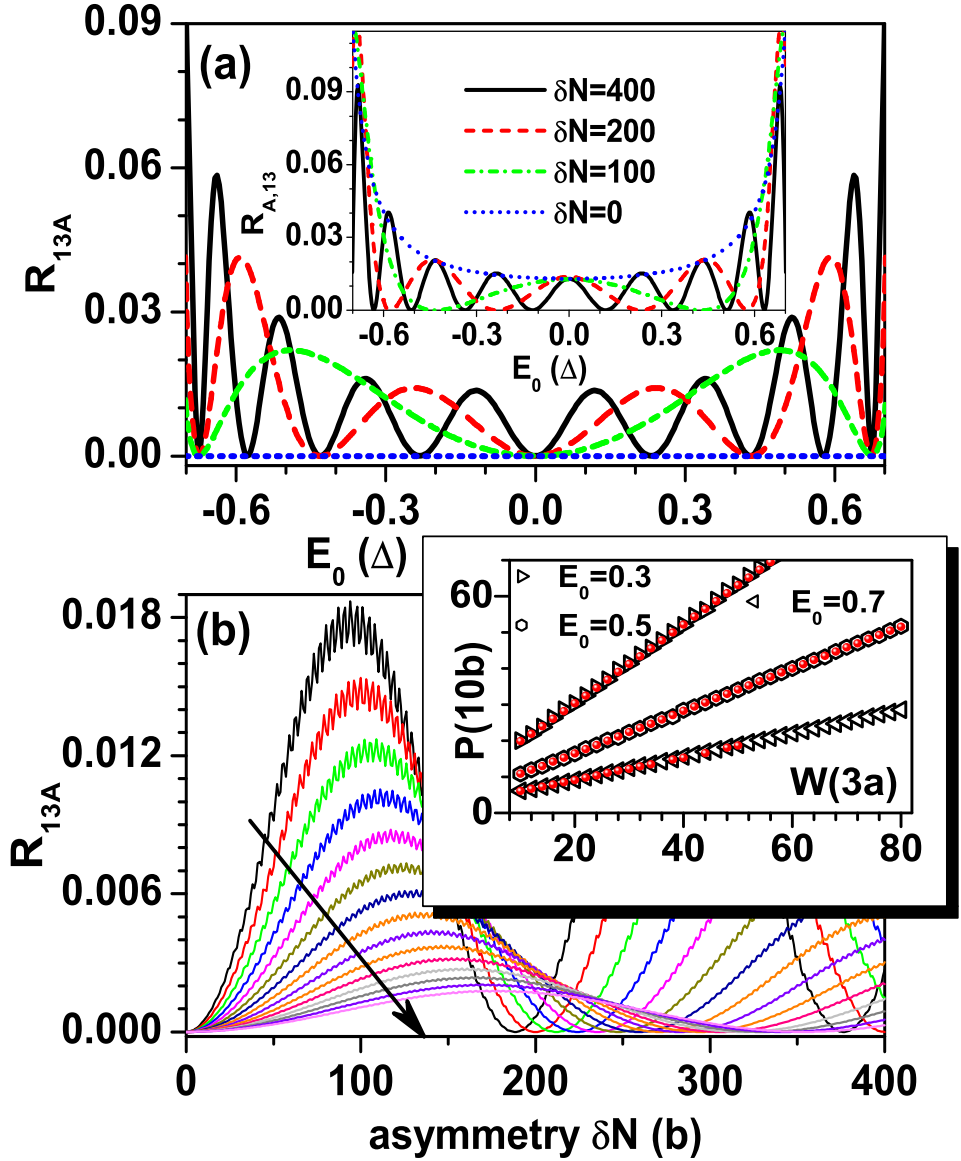}
\caption{ (Color online) Panel (a): With fixed Fermi level
$E_F=0.8$, total AR probability $R_{13A}$ vs Dirac point $E_0$ for
different asymmetry $\delta N$. In the main panel $\delta\varphi=0$,
and while $\delta\varphi=\pi$ in the inset. Panel (b): $R_{13A}$ vs
asymmetry $\delta N$ with $E_0=0.3t$ for different width $W$ from
$10\times 3a$ to $38\times 3a$ with the interval $2\times 3a$ along
the black arrow. Inset panel: the hollow signs are the period $P$
obtained from the main panel and the solid red circles are the
period $P$ from the energy band with the expression
$P=2\pi/(k_x-k_x')$. The other parameters: $\delta\varphi=0$,
$E_F=0.8$.} \label{Fig3}
\end{figure}

\end{document}